\documentclass[pre,twocolumn,showpacs,floatfix,amsmath,amssymb]{revtex4-2}

\usepackage{verbatim}
\usepackage{graphicx,color}

\newcommand{\figOne}{
\begin{figure}[t]
    \centering
        \includegraphics[width=3.0in]{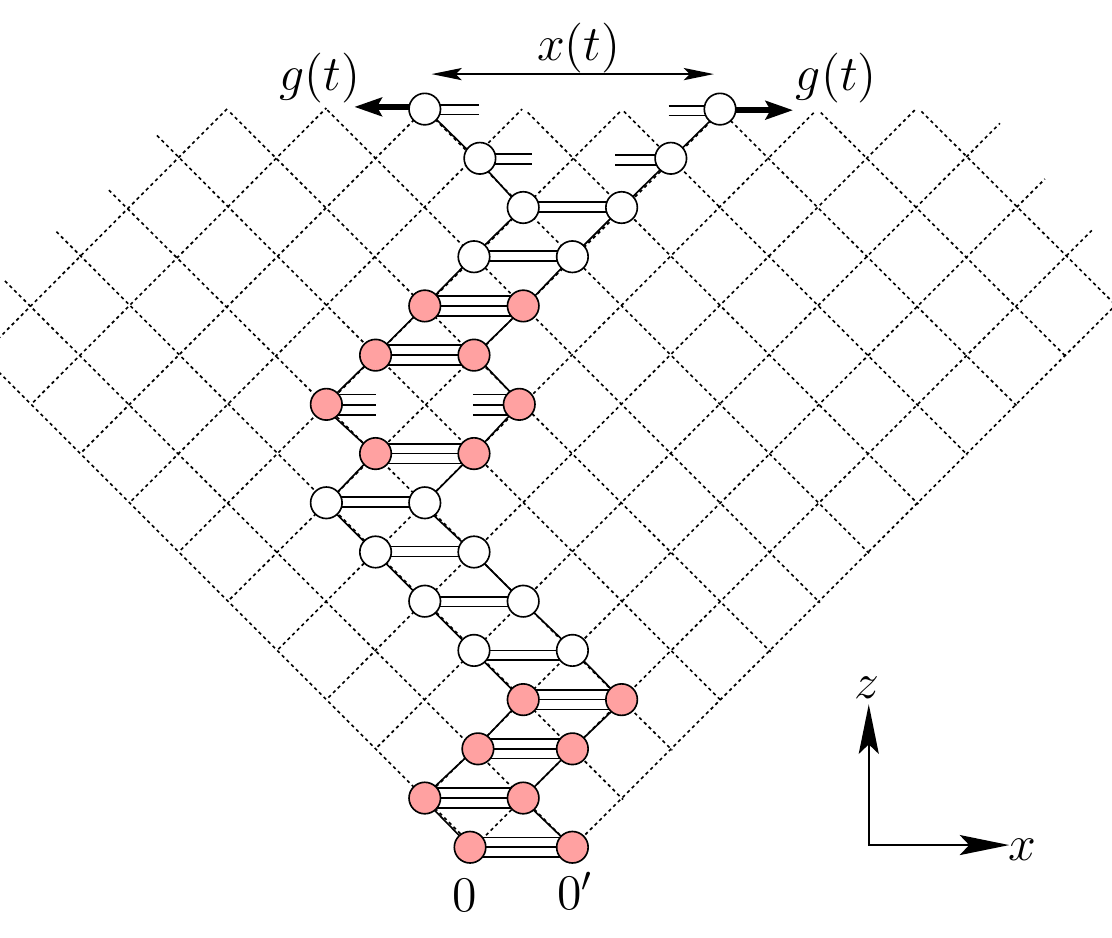}
    
	\caption {Schematic diagram of a heterogeneous dsDNA of the type
	$(B_4A_4)_2$, where $A$ and $B$ represents base pairs having three
	and two hydrogen bonds, respectively. One end of the DNA is anchored
	at the origin ($O$ and $O^{\prime}$), and the strands on the free end
	are subjected to a time-dependent periodic force with frequency
	$\omega$ and amplitude $g_0$. \label{fig:1}}

\end{figure}
}

\newcommand{\figTwo}{
\begin{figure*}[t]
\centering
\includegraphics[width=0.9\linewidth]{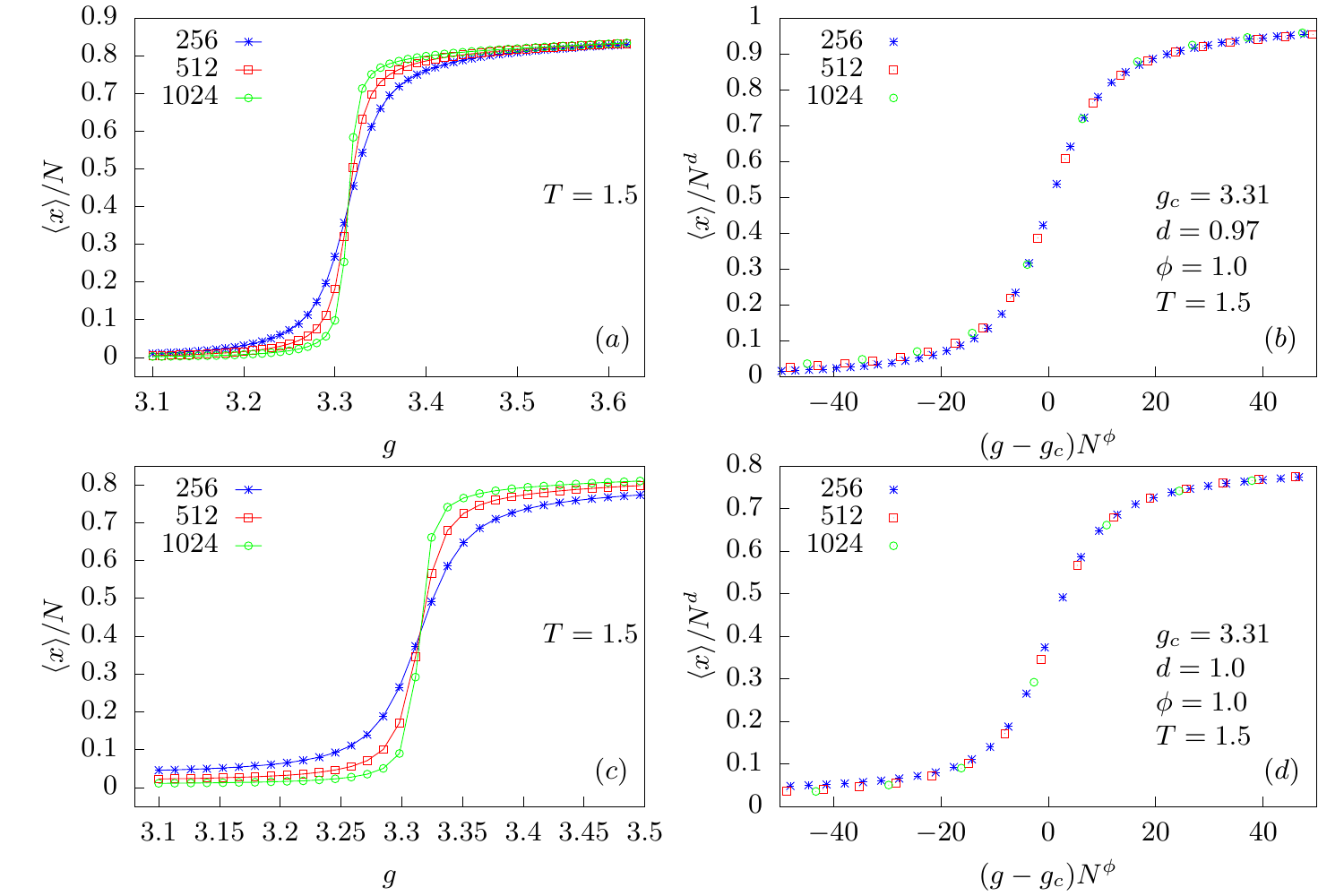}

\caption{Scaled extension $\langle x \rangle / N$, as a function of
	constant pulling force $g$, obtained using the exact transfer matrix
	approach, for different chain lengths $N = 256, 512$, and $1024$ at
	$T = 1.5$ for (a) the heterogeneous sequence $(A_{16}B_{16})_M$. (b)
	$\langle x \rangle / N^d$ as a function of $(g-g_c)N^\phi$ showing a
	nice collapse of data for $g_c = 3.31\pm0.05$, $d = 0.97 \pm 0.05$,
	and $\phi = 1.0\pm0.02$. (c) For the heterogeneous sequence
	$(B_{16}A_{16})_M$. (d) Collapse of data shown in (c) for $g_c =
	3.31\pm 0.05$, $d = 1.0 \pm 0.05$, and $\phi = 1.0\pm 0.02$.  The
	line joining the data points in plots (a) and (c) is just a guide
	for the eye.} \label{fig:2}

\end{figure*}
}

\newcommand{\figThree}{
\begin{figure}[t]
\centering
\includegraphics[width=0.9\linewidth]{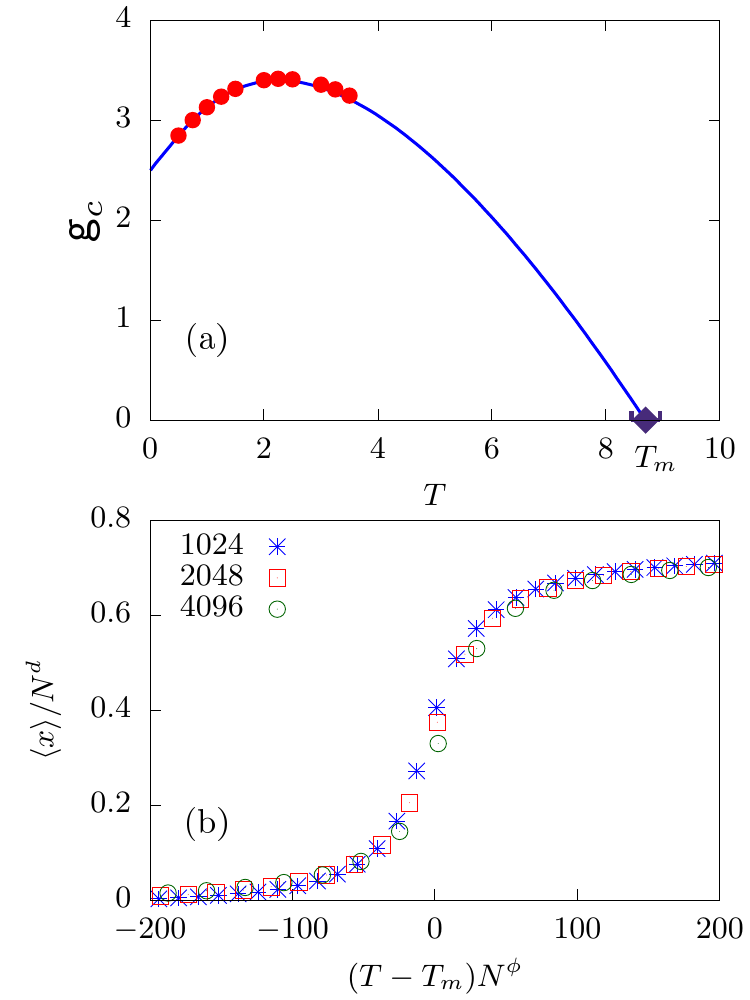}

\caption{(a) Critical unzipping force $g_c$ as a function of temperature
	$T$ for the heterogeneous sequence  $(A_{16}B_{16})_M$.  The line is
	the exact result obtained from the generating function approach
	[Eq.~(\ref{eq:gcff})], and the points are obtained by using
	finite-size scaling of the force-distance isotherms
	[Eq.~(\ref{eq:avx})] as obtained from the exact transfer matrix
	approach. (b) Data collapse of the average distance,  $\langle x
	\rangle $, of the heterogeneous sequence $(A_{16}B_{16})_M$ for $N =
	1024, 2048$, and $4096$. The exponents are $d = 0.52\pm 0.02$, $\phi
	= 0.48\pm0.02$ with melting temperature $T_m = 8.45\pm0.25$.}
	\label{fig:3}

\end{figure}
}

\newcommand{\figFour}{
\begin{figure*}[t]
\centering
\includegraphics[width=0.9\linewidth]{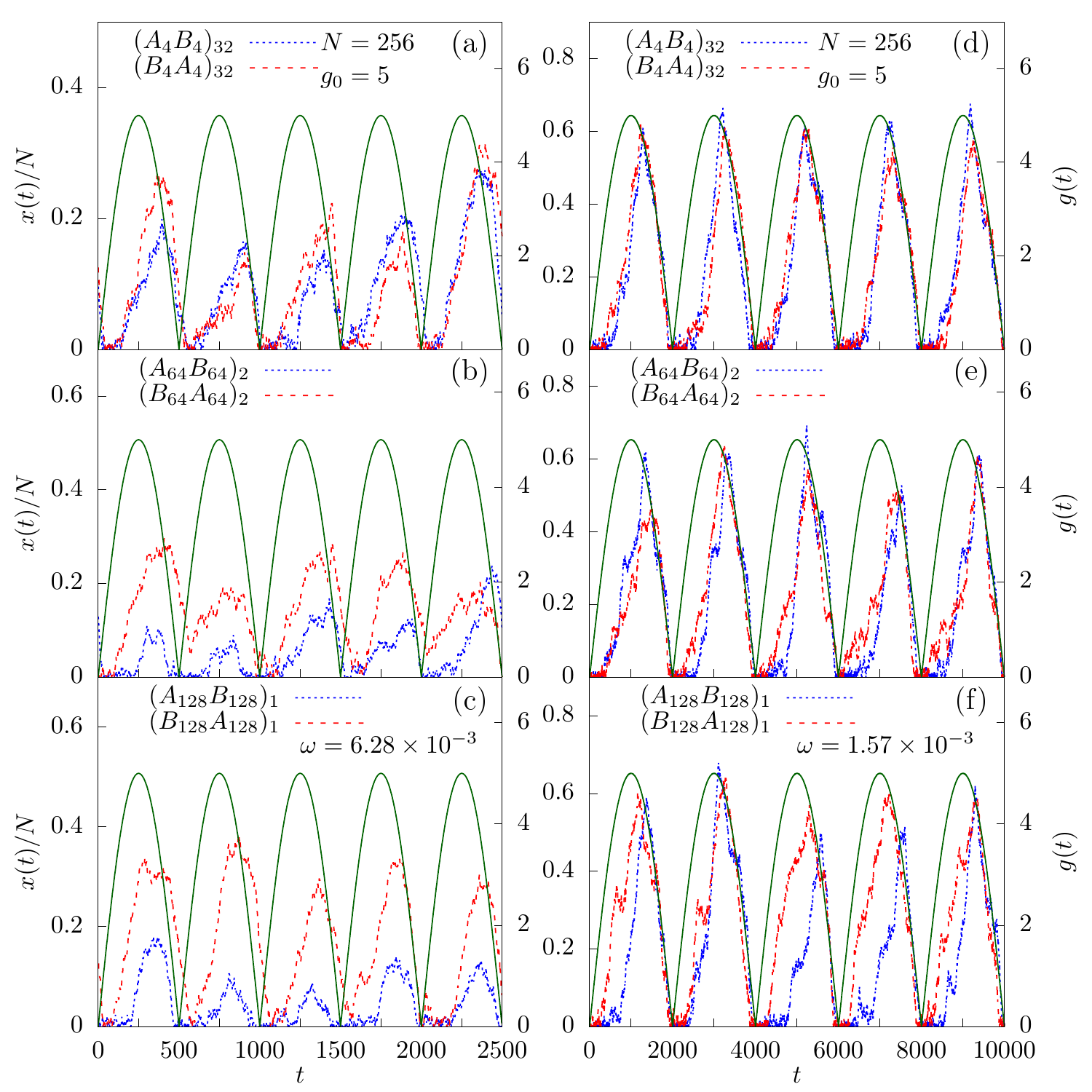}

\caption{ The extension $x(t)$ between the end monomers of the two
	strands of the block copolymer DNA of length $N = 256$ as a function
	of time $t$ when it is subjected to a periodic force of amplitude
	$g_0 = 5$ at frequency $\omega = 6.28\times 10^{-3}$. For the
	sequences (a) $(A_4B_4)_{32}$ and $(B_4A_4)_{32}$, (b)
	$(A_{64}B_{64})_{2}$ and $(B_{64}A_{64})_{2}$, and (c)
	$(A_{128}B_{128})_1$ and $(B_{128}A_{128})_1 $. Plots (d), (e), and
	(f) are same as plots (a), (b), and (c) at frequency $\omega =
	1.57\times 10^{-3}$. The variation of force with time, $g(t)$, is
	represented by solid lines.} \label{fig:4}

\end{figure*}
}

\newcommand{\figFive}{
\begin{figure*}[t]
\centering
\includegraphics[width=0.95\linewidth]{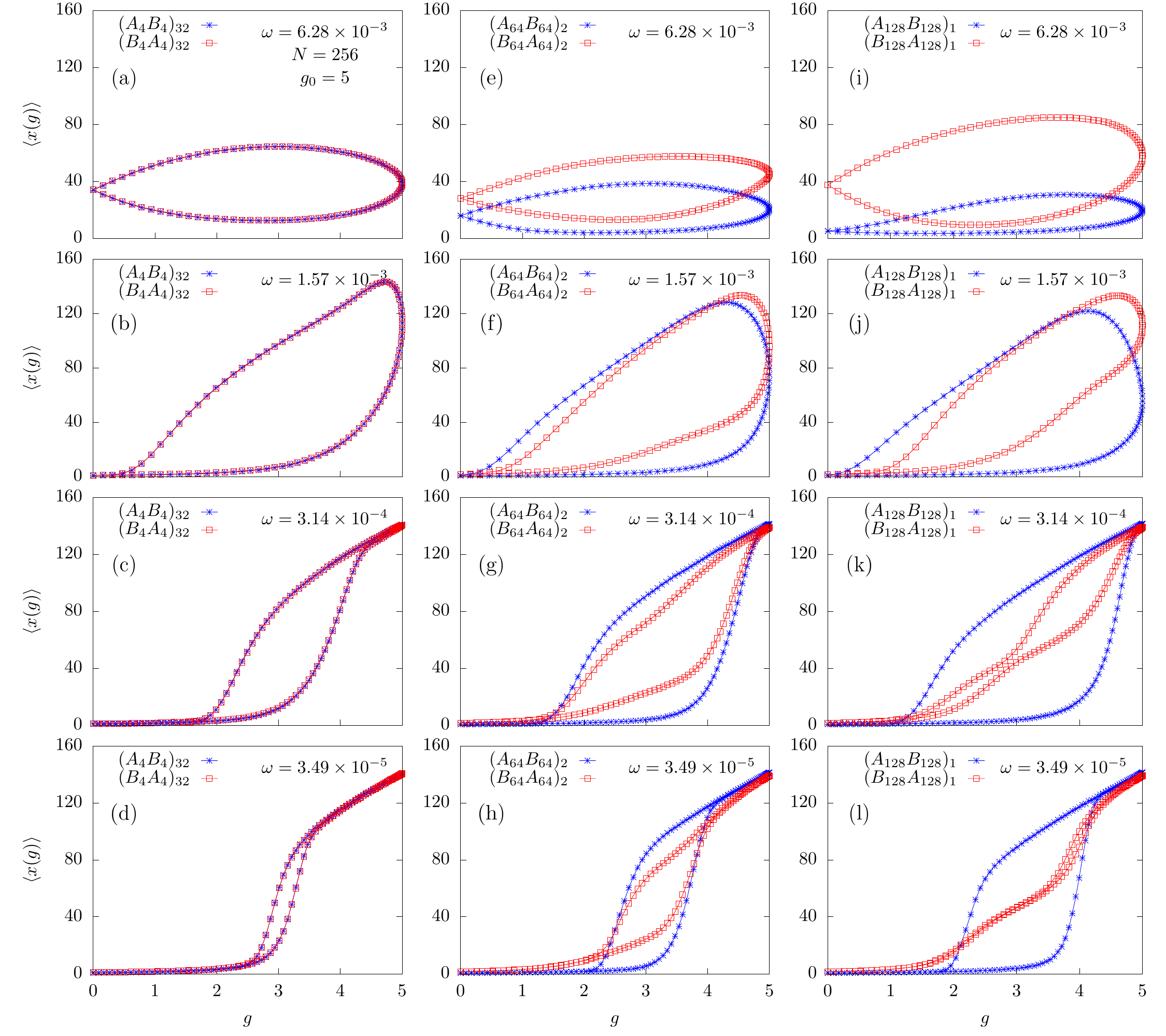}

\caption{ The force $g$ vs extension $\langle x(g) \rangle$ curves
	averaged over $10^4$ cycles for the block copolymer DNA of length $N
	= 256$ and block sizes 8 (first column), 128 (second column), and
	256 (third column) at frequencies $\omega = 6.28 \times 10^{-3}$
	(first row), $\omega = 1.57 \times 10^{-3}$ (second row), $\omega =
	3.14\times 10^{-4}$ (third row), and $\omega = 3.49\times 10^{-5}$
	(fourth row) at force amplitude $g_0 = 5$ and temperature $T = 4$.
	The data shown in this plot are obtained using Monte Carlo
	simulations. The line joining the points is just a guide for the
	eye. } \label{fig:5}

\end{figure*}
}

\newcommand{\figSix}{
\begin{figure*}[t]
\centering

\includegraphics[width=0.95\linewidth]{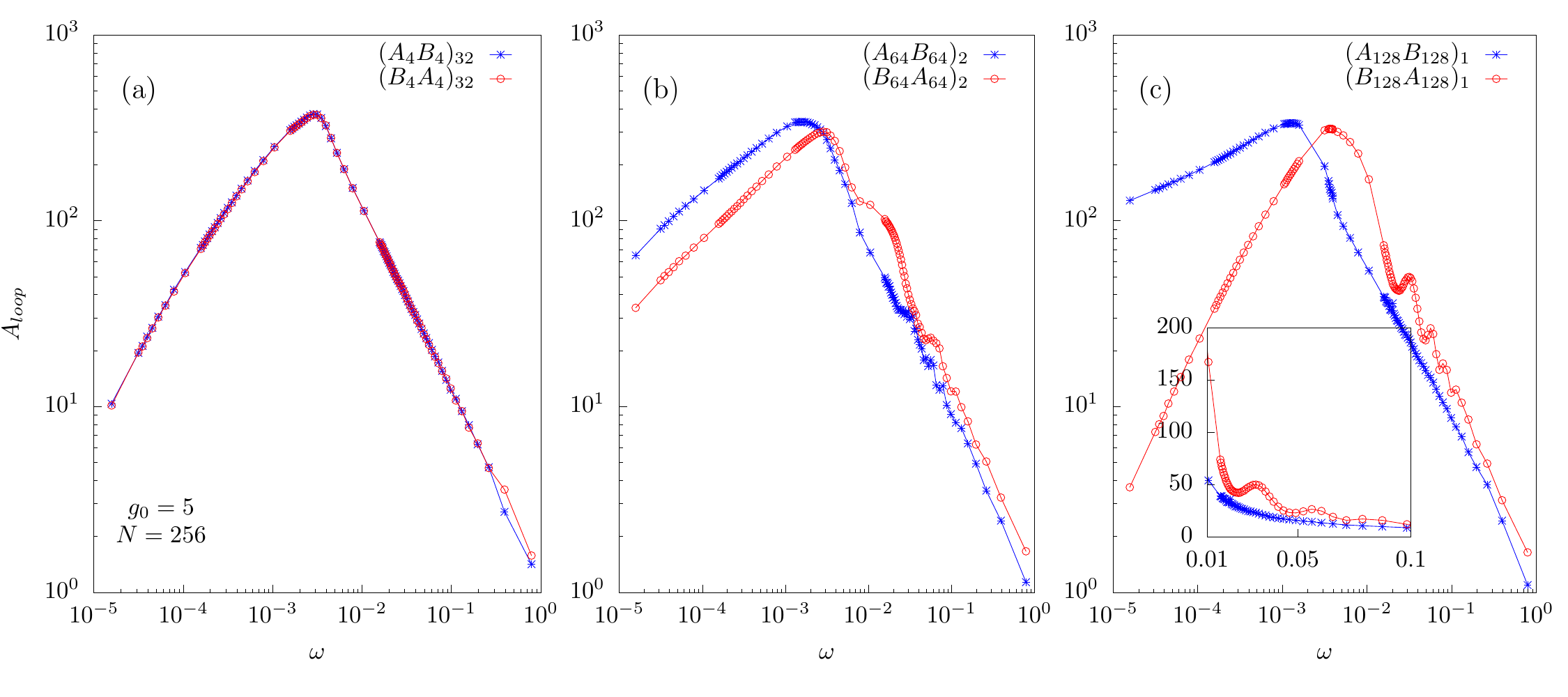}

\caption{ Area of the hysteresis loop $A_{loop}$ as a function of
	frequency $\omega$ (in log-log scale) at force amplitude $g_0 = 5$
	and length $N = 256$ for the block copolymer DNA sequences (a)
	$(A_4B_4)_{32}$ and $(B_4A_4)_{32}$, (b) $(A_{64}B_{64})_{2}$ and
	$(B_{64}A_{64})_{2}$, and (c) $(A_{128}B_{128})_{1}$ and
	$(B_{128}A_{128})_{1}$. The inset shows the higher frequency region
	(in linear scale), where secondary peaks are visible in $A_{loop}$
	for the sequence $(B_{128}A_{128})_{1}$ but are absent for the
	opposite sequence. In all the plots, the line joining the points is
	just a guide for the eye. } \label{fig:6}

\end{figure*}
}

\newcommand{\figSeven}{
\begin{figure*}[t]
\centering

\includegraphics[width=0.85\linewidth]{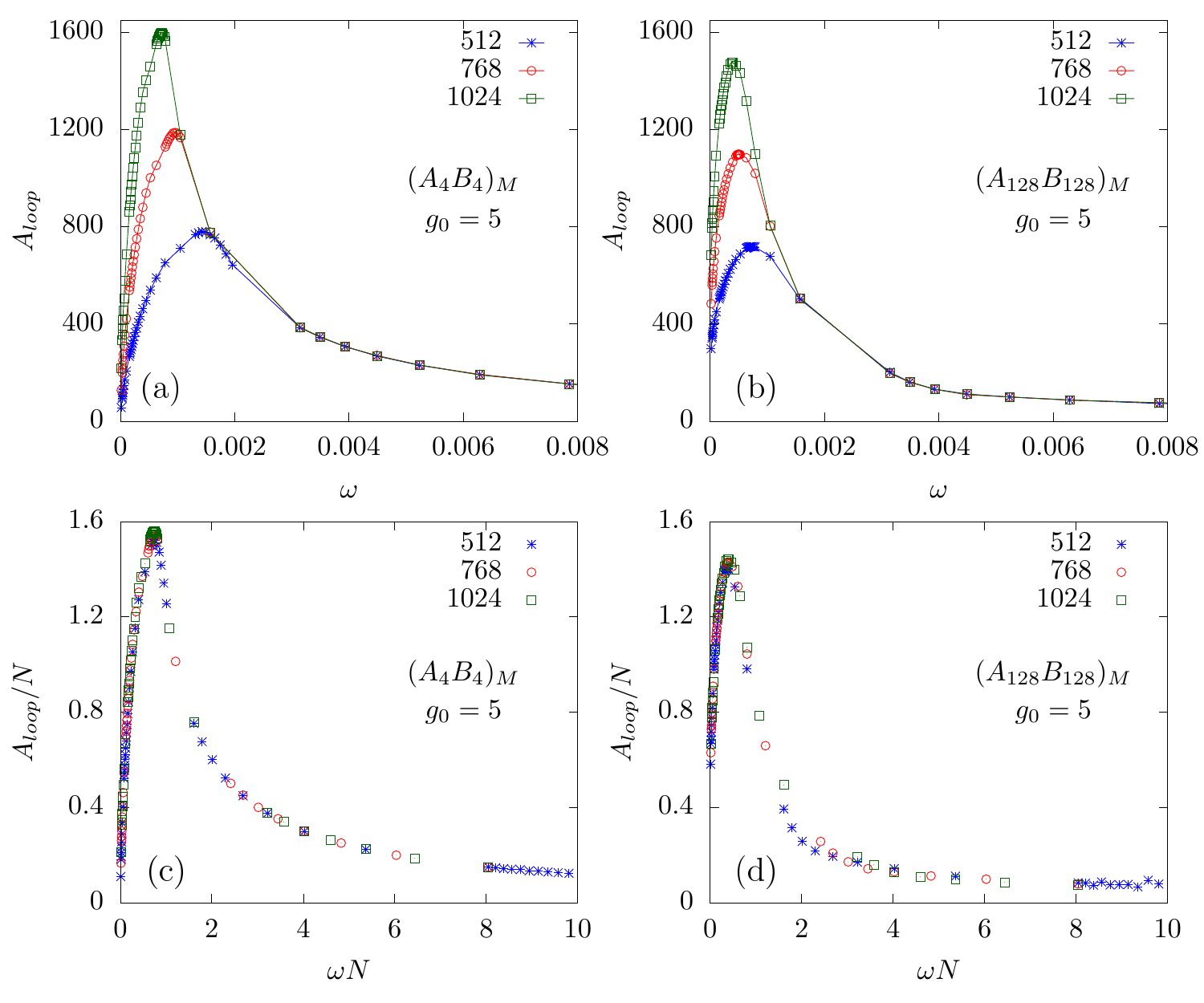}

\caption{Area of the hysteresis loop $A_{loop}$ as a function of
	frequency $\omega$ for the block copolymer DNA of lengths $N = 512,
	768, 1024$ at force amplitude $g_0 = 5$ for sequences (a)
	$(A_4B_4)_M$ and (b) $(A_{128}B_{128})_M$. Plots (c) and (d) are
	$A_{loop}/N$ vs $\omega N$ for respective sequences in (a) and (b).
	The line joining the points in these plots is just a guide for the
	eye. } \label{fig:7}

\end{figure*}
}

\newcommand{\figEight}{
\begin{figure}[t]
\centering
\includegraphics[width=0.9\linewidth]{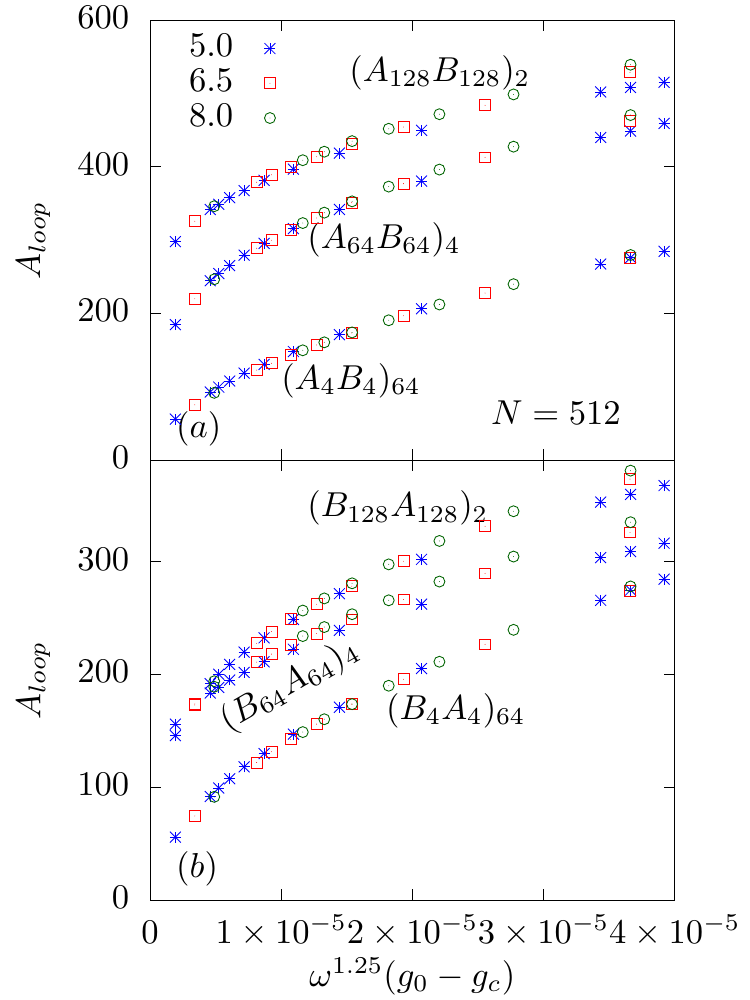}

\caption{ Scaling of $A_{loop}$ with respect to $\omega^{1.25}(g_0-g_c)$
in the low-frequency regime at force amplitudes $g_0 = 5.0, 6.5, 8.0$ of
a block copolymer DNA of length $N = 512$ for the sequences (a)
$(A_4B_4)_{64}$, $(A_{64}B_{64})_{4}$, and $(A_{128}B_{128})_{2}$ and (b)
$(B_4A_4)_{64}$, $(B_{64}A_{64})_{4}$, and $(B_{128}A_{128})_{2}$.}
\label{fig:8}

\end{figure}
}

\begin{document}

\title{Unzipping of a double-stranded block copolymer DNA by a periodic force}

\author{Ramu Kumar Yadav}
\email {ramukumar@iisermohali.ac.in}
\author{Rajeev Kapri}
\email{rkapri@iisermohali.ac.in}
\affiliation{Department of Physical Sciences, Indian Institute of Science Education and
Research Mohali, Sector 81, Knowledge City, S. A. S. Nagar, Manauli PO 140306, India.}

\begin{abstract}
    
	Using Monte Carlo simulations, we study the hysteresis in unzipping
	of a double stranded block copolymer DNA with $-A_n B_n-$ repeat
	units. Here $A$ and $B$ represent two different types of base pairs
	having two- and three-bonds, respectively, and $2n$ represents the
	number of such base pairs in a unit. The end of the DNA are
	subjected to a time-dependent periodic force with frequency
	($\omega$) and amplitude ($g_0$) keeping the other end fixed. We
	find that the equilibrium force-temperature phase diagram for the
	static force is independent of the DNA sequence. For a periodic
	force case, the results are found to be dependent on the block
	copolymer DNA sequence and also on the base pair type on which the
	periodic force is acting. We observe hysteresis loops of various
	shapes and sizes and obtain the scaling of loop area both at low-
	and high-frequency regimes.

\end{abstract}

\date{21 January 2021}

\maketitle

\section{Introduction}

Single-molecule manipulation techniques, which are now used routinely to
study individual molecule by applying mechanical forces in the
pico-newton ranges, has greatly increased our understanding of molecular
interactions in  biological molecules~\cite{Ritort2006}. The unzipping
of a double stranded DNA  (dsDNA) by an external force, exerted by
different enzymes or molecular motors \emph{in vivo}, has biological
relevance in processes like DNA replication and  RNA
transcription~\cite{Watson2003}. The unzipping transition has been
studied for over two decades, both theoretically
\cite{Bhattacharjee2000,Lubensky2000,Sebastian2000,Marenduzzo2001,Marenduzzo2002,Kapri2004}
and experimentally~\cite{Bockelmann2002,Danilowicz2003,Danilowicz2004},
by applying an external pulling force on the strands of the DNA. The
dsDNA unzips to two single strands abruptly when the force exceeds a
critical value showing first order nature of the phase transition
\cite{Bhattacharjee2000,Lubensky2000,Marenduzzo2001,Marenduzzo2002,Kapri2004}.
If biomolecules are subjected to a periodic forcing, they can unbind and
rebind with a hysteresis in their force-distance isotherms. The study of
hysteresis in unbinding and rebinding of biomolecules can provide useful
information on the kinetics of conformational transformations, the
potential energy landscape, controlling the folding pathway of a single
molecule, and in force sensor
studies\cite{Hatch2007,Friddle2008,Tshiprut2009,Li2007,Yasunaga2019}.

In recent years, the behavior of a dsDNA under a periodic force has been
studied using Langevin dynamics simulation of an off-lattice
coarse-grained model for a short homo-polymer DNA
chains~\cite{Kumar2013,Mishra2013,Mishra2013jcp,Kumar2016,Pal2018} and
Monte Carlo simulations on a relatively longer chains of directed
self-avoiding walk (DSAW) model of a homo-polymer dsDNA on a
lattice~\cite{Kapri2012,Kapri2014,Kalyan2019}. In both type of studies,
a dynamical phase transition was found to exist, where the DNA can be
taken from the zipped state to an unzipped state with an intermediate
dynamic state. It was found that the area of the hysteresis loop,
$A_{loop}$, which represents the energy dissipated in the system,
depends on the frequency of the periodic force. At higher frequencies,
it decays with frequency as $A_{loop} \sim 1/\omega$, whereas at lower
frequencies, it scales with the amplitude $g_0$ and frequency $\omega$
of the oscillating force as $A_{loop} \sim g_0^{\alpha} \omega^{\beta}$.
The values of exponents $\alpha$ and $\beta$ are however found to be
different in these studies.

In this paper, we consider a hetero-polymer DNA as a block copolymer
DNA, in which the heterogeneity is considered in the form of repeated
blocks, $A_nB_n$ or $B_nA_n$, where $2n$ is the block length, $A$ and
$B$ are different types of base pairs with two- and three- hydrogen
bonds, respectively. One end of this DNA sequence is subjected to a
pulling force while the other end is kept anchored. We considered both
the constant and the periodic pulling force cases. The unzipping of a
block copolymer DNA by a constant pulling force is found to be a
first-order phase transition. The equilibrium phase boundary separating
the zipped and the unzipped phases does not depend on the DNA sequence
and is found to follow the same exact expression, as obtained for the
homo-polymer DNA case~\cite{Marenduzzo2001,Kapri2004,Kapri2012}, but
with a different effective base pair energy. The results for the
unzipping of a block copolymer DNA subjected to a periodic force are
however found to be sequence dependent. For sequences of higher block
lengths, the results also depend on whether the periodic force is acting
on $A$ type or $B$ type base-pairs. 

The paper is organized as follows: In Sec.~\ref{sec:Model} we define our
model and simulation details. We also define the quantities of interest
we are studying in this paper. Section~\ref{sec:Results} is devoted for
discussions on our results for both the static and the periodic pulling
force cases. We finally summarize the results of this paper in
Sec.~\ref{sec:Summary}. 

\section{Model} {\label{sec:Model} }

\figOne

We define hetero-polymer DNA as a block copolymer DNA of type $(A_n
B_n)_M$, where $A$ and $B$ are two different types of base pairs, $2n$
is the total number of base pairs in a block unit, also be called as
block length, and $M = N/2n$ represents the total number of blocks in
the DNA of length $N$. We consider block lengths $2n = 4, 8, 16, 32, 64,
128$ and $256$. The two strands of the DNA are represented by two
directed self-avoiding random walks on a $(d = 1+1)$ dimensional square
lattice. The walks starting from the origins $O$ and $O^{\prime}$, which
are unit distance apart, are restricted to go towards the positive
direction of the diagonal axis ($z$ direction) without crossing each
other. The directional nature of the walks take care of the
self-avoidance. Whenever the complementary bases are unit distance apart
they gain energy of $-2\epsilon$ ($\epsilon >0$) for the base pair of
type $A$ and $-3\epsilon$ for the base pair of type $B$. Here we have
assumed that $\epsilon$ ($\epsilon >0$), represents the strength of a
hydrogen bond. 

Two strands of the DNA at one end are always kept fixed at origins $O$
and $O^{\prime}$ and the other end monomers are subjected to a
time-dependent periodic force $g(t)$
\begin{equation}                 
     g(t) = g_0 \mid\sin(\omega\*t)\mid, 
     \label{eq:force}  
\end{equation}        
where $g_0$ is the amplitude and $\omega$ is the frequency. The
schematic diagram of the model is shown in Fig.~\ref{fig:1}.

In this paper we consider the following two cases: (i) the base pairs
having two hydrogen bonds are anchored at the origins and the time
varying force is applied on the base pairs that are bound by three
hydrogen bonds, represented by $(A_nB_n)_{M}$, and (ii) the opposite
case, i.e., the base pairs having three hydrogen bonds are anchored at
the origins and the force is acting on monomers that are bound by two
hydrogen bonds (represented by $(B_n A_n)_{M}$). While the equilibrium
results for both the cases are found to be the same, the nonequilibrium
results show marked differences.

We perform Monte Carlo simulations of the model using the Metropolis
algorithm. The strands of the DNA undergo Rouse dynamics that consists
of local corner-flip or end flip moves that do not violate mutual
avoidance~\cite{Doi1986}. The elementary move consists of selecting a
random monomer from a strand, which itself is chosen at random, and
flipping it. If move results in the overlapping of two complementary
monomers, thus forming a base-pair between the strands, it is always
accepted as a move. The opposite move (i.e., unbinding of monomers) is
chosen with the Boltzmann probabilities $\eta=\exp{(-2\epsilon/k_B T)}$
or $\eta=\exp{(-3\epsilon / k_B T)}$ for base pairs of types $A$ and
$B$, respectively. If the chosen monomer is unbind, which remains unbind
after the move is performed is always accepted. The time is measured in
units of Monte Carlo steps (MCSs). One MCS consists of $2N$ flip
attempts, which means that on average, every monomer is given a chance
to flip. Throughout the simulation, the detailed balance is always
satisfied and the algorithm is ergodic in nature. It is always possible,
from any starting DNA configuration, to reach any other configuration by
using the above moves. We let the simulation run for $2000\pi/\omega$
MCSs, so that system reaches the stationary state before taking
measurements. Throughout this paper, we have chosen dimensionless
quantities. The quantities having dimensions of energy are measured in
units of $\epsilon$ and the quantities that have dimensions of length
are measured in terms of the lattice constant $a$. In this paper we have
taken $k_B = 1$, $\epsilon = 1$, and $a = 1$. 

The separation between the end monomers of the two strands, $x(t)$,
changes under the influence of the applied external force $g(t)$, is
monitored as a function of time $t$. The time averaging of $x(t)$ over a
complete period 
\begin{equation}
   Q = \frac{\omega} {\pi}\*\oint x(t)\*dt 
       \label{fig:OrderPara}
\end{equation}
can be used as a dynamical order parameter\cite{Chakrabarti1999}. From
the time series $x(t)$, we obtain the extension $x(g)$ as a function of
force $g$ and average it over 10000 cycles to obtain the average
extension $\langle x(g) \rangle$ as a function of $g$. For systems far
away from equilibrium, the average extension, $\langle x(g) \rangle$,
for the forward and backward paths for the periodic force is not the
same, and we see a hysteresis loop. The area of hysteresis loop,
$A_{loop}$, is defined by 
\begin{equation}
     A_{loop} = \oint \langle x(g) \rangle dg  
     \label{eq:A_loop}
\end{equation} 
depends upon the frequency $\omega$ and the amplitude $g_0$ of the
oscillating force. This quantity also serves as another dynamical order
parameter.

\section{Results and Discussions} {\label{sec:Results} }

In this section we discuss the results obtained for both the static and
the dynamic cases. Let us first take the static case.

\figTwo

\subsection{Static Case ($\boldsymbol{\omega = 0)}$}

\figThree

In the static case, this model can be solved exactly using the
generating function and the exact transfer matrix techniques. If the
partition function of the dsDNA of length $n$ with separation $x$
between monomers of the strands is represented by $\mathcal{D}_n(x)$, in
the fixed distance ensemble, then $\mathcal{D}_n(x)$ satisfies the
recursion relation:
\begin{equation}    \label{eq:recrel}
    \mathcal{D}_{n+1}(x) = \left[ \mathcal{D}_n(x+1) + 2\mathcal{D}_n(x)
	+ \mathcal{D}_n(x-1) \right] \times \mathcal{C},
\end{equation}
where
\begin{equation}    \label{eq:C}    
    \mathcal{C}  = \begin{cases}
    1 + \left( e^{2 \beta \epsilon} - 1\right) \delta_{x,1}, & \text{for
	base pair type $A$} \\
    1 + \left( e^{3 \beta \epsilon} - 1\right) \delta_{x,1}, & \text{for
	base pair type $B$}.
    \end{cases}
\end{equation}
The above recursion relation can be iterated $N$ times, with an initial
condition $\mathcal{D}_0(x)=\delta_{x, 1}$ to obtain the partition
function of the DNA of length $N$. The recursion relation
(Eq.~(\ref{eq:recrel})) with a single base pairing energy (say
$\varepsilon$) for each base pair such that $\mathcal{C} = \left[ 1 +
\left( e^{\beta \varepsilon} - 1\right) \delta_{x,1} \right]$ has been
solved exactly via the generating function technique
\cite{Marenduzzo2001,Marenduzzo2002,Kapri2004} to obtain the exact
unzipping phase diagram. In this method, the singularities of the
generating function are calculated. The phase of the DNA is given by the
singularity closest to the origin and when the two singularities cross
each other a phase transition takes place. Taking the following form for
the generating function for $\mathcal{D}_n(x)$,
\begin{equation}
    \hat{\mathcal{D}}(z,x) = \sum_n z^n \mathcal{D}_n(x) = \kappa^{x}(z)
	Y(z),
\end{equation}
and used in the above recursion relation (Eq.(\ref{eq:recrel}) with
initial condition $\mathcal{D}_0(x) = \delta_{x,1}$), we obtain
$\kappa(z) = (1- 2z -\sqrt{1-4z})/(2z)$ and $Y(z) = 1/ \left[ 1 - z
\left( 2 + \kappa(z) \right) e^{\beta \varepsilon} \right]$. The
singularities of $\kappa(z)$ and $Y(z)$ are $1/2$ and $z_2 = \sqrt{1 -
e^{-\beta \varepsilon} } - 1 + e^{-\beta \varepsilon}$, respectively. The
zero force melting, which comes from $z_1 = z_2$, takes place at a
temperature  $T_m = \varepsilon/\ln (4/3)$. In the large length limit,
$D_n(x)$ can be approximated as $D_N(x) \approx \kappa^x(z_2) /
z_2^{N+1}$, with the free energy $\beta F = N \ln z_2 - x \ln
\kappa(z_2)$. The average force required to maintain the separation $x$,
in a fixed distance ensemble, is then given by
\begin{equation}
    g(T) = \frac{\partial F}{\partial x} = - k_B T \ln \kappa (z_2).
    \label{eq:gc}
\end{equation}
In the fixed force ensemble, the generating function can be written as
\begin{eqnarray}
    \mathcal{G}(z, \beta, g_0) &=& \sum_x e^{2\beta g_0 x} \sum_{n} z^n
	\mathcal{D}_n(x) =  \sum_x e^{2 \beta g_0 x} \kappa^x(z) Y(z) \cr
    &=& \frac{Y(z)}{1 - \kappa(z) e^{2\beta g_0}},
\end{eqnarray}
which has an additional force-dependent singularity $z_3 = 1/[2 + 2
\cosh (2\beta g_0)]$. The phase boundary comes from $z_2=z_3$, and is
given by
\begin{equation}\label{eq:gcff}
    g_c(T) = k_B T \cosh^{-1} \left [ \frac{1}{2} \frac{1}{ \sqrt{1
	-e^{-\beta \varepsilon} } - 1 + e^{-\beta \varepsilon}} -1 \right ],
\end{equation}
which is same as the phase boundary obtained in the fixed distance
ensemble [Eq.~(\ref{eq:gc})]. In the above expression, $\varepsilon$ is
the only free parameter, which can be tuned. For the block copolymer DNA
case, in every block, we have $n$ base pairs each of types $A$ and $B$,
giving the total base pairing energy $(2\epsilon + 3\epsilon)n$. Since
the total energy of the block remains the same irrespective of the
sequence $(A_n B_n)_M$ or $(B_n A_n)_M$, we seek if an effective base
pairing energy $\varepsilon = 5\epsilon/2$ in Eq.~(\ref{eq:gcff}) can
give us the exact phase boundary for the block copolymer DNA as obtained
by iterating the recursion relation Eq.~(\ref{eq:recrel}). The phase
diagram of unzipping of a block copolymer DNA (with $\varepsilon =
5\epsilon/2$) is shown in Fig.~\ref{fig:3}(a) by solid lines. 

The exact transfer matrix technique can be used to obtain many other
equilibrium properties which are based on thermal averaging for a finite
system size. In this technique, the partition function
$\mathcal{D}_N(x)$ for the DNA of length $N$, at any temperature, can be
obtained numerically by iterating the above recursion relation [i.e.
Eq.(\ref{eq:recrel})] $N$ times, with an initial condition
$\mathcal{D}_0(1) = 1$. The equilibrium average separation between the
end monomers, $\langle x \rangle_{\rm eq}$, can then be obtained by
\begin{equation}\label{eq:avx}
    \langle x \rangle_{\rm eq} = \frac{ \sum_x x \ \mathcal{D}_N(x)
	e^{\beta g_0 x}}{ \sum_x \mathcal{D}_N(x) e^{\beta g_0 x} }.
\end{equation}

\figFour

In Fig.~\ref{fig:2}, we have plotted the scaled extension $\langle x
\rangle / N$, as a function of constant pulling force $g$ for different
chain lengths $N = 256, 512$, and $1024$ at $T = 1.5$ obtained by
iterating the recursion relation Eq.~(\ref{eq:recrel}) for the
heterogeneous sequences $(A_{16}B_{16})_M$ [Fig.~\ref{fig:2}(a)], in
which the base pair of type $A$ is anchored at the origin and an
external force $g$ is applied on the base pair type $B$, and
$(B_{16}A_{16})_M$ [Fig.~\ref{fig:2}(c)], which is the opposite of the
above. From the figure, we can clearly see that the DNA is in the zipped
phase at lower $g$ values and in the unzipped phase when $g$ exceeds a
critical value $g_c$. Furthermore, as the length $N$ of DNA increases,
the transition becomes sharper. In the thermodynamic limit, it would
become a step function at a critical value $g_c$. The point of
intersection of these isotherms for various lengths is very close to
the critical force $g_c$. We use the finite-size scaling, shown in
Figs.~\ref{fig:2}(b) and 2(d), to extract the value of critical force
$g_c$. The critical force, $g_c = 3.31\pm0.05$ (at $T = 1.5$), is found to
be same for the both sequences $(A_{16}B_{16})_M$ and $(B_{16}A_{16})_M$
implying that, at equilibrium, it does not matter whether the DNA is
unzipped from the end having base pairing with three hydrogen bonds
(stronger) or the base pairing with two hydrogen bonds. This is because
the unzipping transition is a first-order phase transition. The critical
forces obtained at various temperatures using the transfer matrix method are
shown in Fig.~\ref{fig:3}(a) by  points. They match exactly with the
analytical results given by Eq.~(\ref{eq:gc}). The same exact transfer
matrix technique could also be used to obtain the melting temperature of
the DNA. We again iterate the recursion relations now at zero force
value $g = 0$ and obtain the equilibrium separation between strands at
the free end as a function of temperature. We use chain lengths $N =
1024, 2048$, and $4096$, and the finite-size scaling of the form 
\begin{equation}
\langle x \rangle  = N^d \mathcal{G}\Big( \big(T-T_m\big) N^\phi \Big) ,
\label{eq:tc}
\end{equation}
to obtain the melting temperature $T_m$. A nice collapse is obtained for
$d = 0.52 \pm 0.02$, $\phi = 0.48 \pm 0.02$ and $T_m = 8.45 \pm 0.25$
for sequence$(A_{16}B_{16})_M$ [shown in Fig.~\ref{fig:3}(b)]. We have
tried various other sequences and found that the melting temperatures
for all the heterogeneous sequences allowed in our model are the same.
The melting temperature obtained by the transfer matrix method is also
shown in Fig.~\ref{fig:3}(a) by a diamond.

\subsection{Dynamic Case}

In the previous section, we have seen that the unzipping of a block
copolymer DNA in equilibrium does not depend on whether the force acts
on the base pairs of type $A$ or type $B$. However, for the time-dependent
periodic force, we find that the unzipping depends on which base pairs
are unzipped first.

In Fig.~\ref{fig:4}, we have plotted the time variation of external force
$g(t)$ and scaled extension $x(t)/N$ for the DNA of length $N = 256$
with respect to time $t$ for five consecutive cycles when it is
subjected to a periodic force of amplitude $g_0 = 5$ at two different
frequencies $\omega = 6.28\times 10^{-3}$ and $1.57\times 10^{-3}$ at $T
= 4$. The force increases from zero to a maximum value of $g_0$, which is
much larger than the critical force $g_c$ needed to unzip the DNA at
equilibrium, and then decreases to zero again. The DNA responds to this
external force and starts unzipping slowly. We can see that there is
always a lag between the scaled extension and the force. It is easy to
understand that, for a homopolymer DNA, the time required to unzip a
dsDNA is directly proportional to its length. The larger the length of the
DNA, the more is the unzipping time. However, for a block copolymer DNA, the
unzipping time for the DNA of same length can be quite different as it
also depends on its sequence. Figure \ref{fig:4}(a) shows the time
variation of the distance between end monomers of the two strands for
sequences of smaller block sizes $8$ [$(A_4B_4)_{32}$ and
$(B_4A_4)_{32}$]. The scaled extension for both sequences is almost
the same. However, on increasing the block sizes to $128$ but keeping
the frequency and amplitude same, the scaled extension for the sequence
$(B_{64}A_{64})_2$ is more than that for the opposite sequence
$(A_{64}B_{64})_2$ [Fig.~\ref{fig:4}(b)]. On increasing the block size
further to $256$, the scaled extension for the sequence
$(B_{128}A_{128})_1$ becomes almost double that for the opposite
sequence $(A_{128}B_{128})_1$ as shown in Fig.~\ref{fig:4}(c). This can
be understood as follows. In one cycle of the periodic force with higher
frequency ($\omega = 6.28\times 10^{-3}$), the force changes faster and
the system gets less time to relax. Since it is easier to break base
pairs with two hydrogen bonds (type A) in comparison with base pairs with
three hydrogen bonds (type B), more base pairs are broken for
the sequence $(B_{128}A_{128})_1$ than for the sequence
$(A_{128}B_{128})_1$, and we see the higher extension. However, on lowering
the frequency of the external force to $1.57\times 10^{-3}$, the system
gets enough time to relax, and the extension between the strands become
almost comparable for both the sequences for all block sizes as shown in
Figs.~\ref{fig:4}(d)-4(f).

\subsubsection{Hysteresis loops}

\figFive

\figSix

\figSeven

We have seen that the extension $x(t)$ follows the driving force $g(t)$
with a lag. When it is averaged over various cycles, we obtain the
average extension $\langle x(g) \rangle$ as a function of force $g$
showing a closed loop. The shape of a loop tells much about the
dynamics of the system and depends on the frequency $\omega$ and the
force amplitude $g_0$. For the present problem, the hysteresis loop also
depends on the sequence of the block copolymer DNA. In Fig.~\ref{fig:5},
we have plotted $\langle x(g) \rangle$ as a function of force $g$ at
four different frequencies $\omega = 6.28 \times 10^{-3}$, $1.57 \times
10^{-3}$, $3.14 \times 10^{-4}$, and $3.49 \times 10^{-5}$ at force
amplitude $g_0 = 5$ for the DNA of length $N = 256$ with block sizes 8,
128, and 256 at $T = 4$. All of them show hysteresis loops but with
different shapes. The loops for DNA of smaller block
sizes, e.g., $(A_4B_4)_{32}$ and $(B_4A_4)_{32}$
[Figs.~\ref{fig:5}(a)-5(d)], are almost the same, irrespective of which
base pair is acted upon by the driving force. To understand the shapes
of the loop, we first note that at higher frequency, i.e., $\omega =
6.28 \times 10^{-3}$, the stationary state of the DNA at $g = 0$ is a
partially unzipped state with an average extension $\langle x(g) \rangle
= 35$. At this frequency, the force changes very rapidly and the strands
of the DNA do not get enough time to relax, and only a small loop is
traced by the extension between them. However, for a relatively lower
frequency $\omega = 1.57 \times 10^{-3}$, the stationary state of the
DNA at $g = 0$ is a fully zipped configuration with an average extension
$\langle x(0) \rangle = 0$. The strands now get relatively more time to
relax, and the loop area increases. Even at this frequency, the DNA does
not get fully unzipped at the maximum force value. This is shown by the
rounding of the loop at the maximum force value. The extension increases
even though the force decreases. It reaches a maximum for some lower
force value, in the backward cycle, and then decreases to zero when $g =
0$. On decreasing the frequency further, the isotherms at higher and
lower force values start following the same curve for the forward and
backward cycles but with a loop in between whose area decreases with
decreasing frequency. The situation for the higher block lengths are, however, different. For the sequence $(A_{128}B_{128})_{1}$ (third column
in Fig.~\ref{fig:5}), the stationary state at frequency $\omega = 6.28
\times 10^{-3}$ is a completely zipped configuration with an average
extension $\langle x(g) \rangle \approx 0$ at $g = 0$ [see
Fig.~\ref{fig:5}(i)]. This is because the driving force is acting on
base pairs with three hydrogen bonds having higher strength, and hence
only a few base pairs are broken. Therefore the area of the loop traced by
the extension between the strands is also small. In contrast, for the
sequence $(B_{128}A_{128})_{1}$, the stationary state (at the same
frequency) is a partially unzipped DNA. In this case, the driving force
can break more bonds as it is acting on the base pairs with
two hydrogen bonds and therefore is weaker than the previous case.
Therefore, the average extension $\langle x(g) \rangle$ traces a loop
with larger area. On decreasing the frequency to $\omega = 1.57 \times
10^{-3}$, the stationary state (at $g = 0$) for the sequence
$(B_{128}A_{128})_{1}$ changes to a fully zipped configuration [see
Fig.~\ref{fig:5}(j)] as the strands now get enough time to relax and get
re-zipped again for forces far below the critical value. On decreasing
the frequency further to $\omega = 3.14 \times 10^{-4}$
[Fig.~\ref{fig:5}(k)], the strands get equilibrated for smaller and
larger force values, and therefore the extension starts following the
equilibrium curve at these force values. However, there is still a
hysteresis curve at the transition region that decreases on decreasing
the frequency of the force. It is found that the size of the hysteresis
loop for the sequence $(B_{128}A_{128})_{1}$ decreases much faster. The
shape of the loop for this sequence starts closing at the center, and the
loop divides into two smaller loops and a plateau starts emerging [see
Fig.~\ref{fig:5}(k)]. At frequency $\omega = 3.49 \times 10^{-5}$, one
of the smaller loop completely disappears, and the other loop is also
very small [Fig.~\ref{fig:5}(l)]. The shape of loop is markedly
different for the opposite sequence $(A_{128}B_{128})_{1}$ at the same
frequency with considerable loop area. For sequences of intermediate
block lengths, for example, $(A_{64}B_{64})_2$ and $(B_{64}A_{64})_2$
with block length $128$ [Figs.~\ref{fig:5}(e)-5(h)], the hysteresis
loops show mixed features as seen for sequences with smaller and larger
block lengths (columns one and three of Fig.~\ref{fig:5}).

\subsubsection{Loop area}

We calculate the area of the hysteresis loops, shown in
Fig.~\ref{fig:5}, numerically using the trapezoidal rule. For the trapezoidal
rule to work properly, the intervals should be uniformly spaced. For the
problem considered in this paper, the force increases as sine function
which gives us non uniformly spaced force values. To convert it into
a uniformly spaced interval, we divide the force interval $g \in [0,g_0]$,
for both the rise and fall of the cycle, into 1000 equal intervals, and
then obtain the value of $\langle x(g) \rangle$ at the end points of
these intervals by interpolation using cubic splines of the GNU
Scientific Library~\cite{Galassi2009}. The loop area, $A_{loop}$, is
then evaluated numerically by using the trapezoidal rule on these
intervals.

In Fig.~\ref{fig:6}, the area of the hysteresis loop, $A_{loop}$, is
plotted as a function of the frequency, $\omega$, of the external
pulling force for sequences of various block sizes 8, 128, and 256 of
block copolymer DNA of length $N = 256$ at force amplitude $g_0 = 5$ and
temperature $T = 4$. On decreasing the frequency of the pulling force,
it is found that the loop area first increases, reaches a maximum value
at some frequency $\omega^{*}$, and then decreases with decreasing the
frequency further, similar to the hysteresis loop area behavior for a
homopolymer DNA under periodic forcing ~\cite{Kapri2014}. At a frequency
$\omega^{*}$, the natural frequency of the block copolymer DNA matches
the frequency of the externally applied force, and we have a resonance
with a maximum loop area. For the block copolymer DNA case, the
behaviour of $A_{loop}$ also depends on the DNA sequence used. For
sequences of smaller block lengths, e.g., $(A_4B_4)_{32}$ and
$(B_4A_4)_{32}$, the loop area is same, and hence $\omega^{*}$ is same
for both the sequences [see Fig.~\ref{fig:6}(a)]. However, this is no
longer true for sequences of higher block sizes, where clear differences
are seen for the opposite sequences. For example, for block length
$256$, we can observe that the frequency $\omega^{*}$ is higher for the
sequence $(B_{128}A_{128})_1$ than its opposite sequence
$(A_{128}B_{128})_1$ [see Fig.~\ref{fig:6}(c)]. At frequencies higher
than $\omega^{*}$, the former sequence shows secondary peak structures,
whereas the sequence $(A_{128}B_{128})_1$ falls off smoothly without
showing any such peaks [see inset of Fig.~\ref{fig:6}(c)].  On the lower
side of frequency $\omega^{*}$, the loop area $A_{loop}$ falls sharply
to zero for the sequence $(B_{128}A_{128})_1$, whereas it decreases very
slowly for the opposite sequence $(A_{128}B_{128})_1$.  The secondary
peaks in $A_{loop}$ are the frequencies $\omega_p = (2p -1)\pi/2N$, with
$p = 1,2, \ldots$ as integers, and are higher Rouse
modes~\cite{Kapri2014}. These modes are more pronounced for the sequence
$(B_{128}A_{128})_1$, where the pulling force is applied to $A$ type
base pairs that can be broken at relatively lower force values than $B$
type base pairs, and hence more base pairs are broken. The two strands
thus separated with each other can explore more configurations and can
trace a hysteresis loop [See Fig.~\ref{fig:5}(l].  This loop has larger
area whenever the frequency of the periodic force is $\omega_p$, i.e.,
higher harmonics of the natural frequency of the DNA. In contrast, for
the opposite sequence $(A_{128}B_{128})_1$, more force is required to
break $B$ type of base pairs where force is applied, and at higher
frequencies only a few base pairs are broken, and the loop traced by
unzipped strands is very small and hence no secondary peaks are visible.

We have plotted $A_{loop}$, vs $\omega$ for various sequences at force
amplitude $g_0 = 5$ for block copolymer DNA of three different lengths
$N = 512$, 768, and 1024 in Figs.~\ref{fig:7}(a) and 7(b). The maximum
value of the loop area $A_{loop}$ is directly proportional to the length
of the DNA used in the simulation. Furthermore, the resonance frequency
$\omega^{*}$, where $A_{loop}$ is maximum, decreases with the length of
the DNA, suggesting the scaling form 
\begin{equation}
	A_{loop} = N^{d} \mathcal{F}( \omega N^z),
\end{equation}
where $d$ and $z$ are exponents, for the loop area $A_{loop}$.  When 
$A_{loop}/N$, is plotted with the scaled frequency $\omega N$ (i.e., 
for exponents $d=1$ and $z=1$) , we find a nice data collapse for 
sequences of all the block sizes [Figs.~\ref{fig:7}(c) and 7(d)], 
implying that the loop area for the block copolymer DNA decreases 
with frequency $\omega$ as $A_{loop} \sim 1/\omega$ at higher frequencies 
(i.e., $\omega \to \infty$), similar to the homopolymer
case~\cite{Kapri2014}.

\figEight

To obtain the scaling behavior at lower frequencies, we have plotted
$A_{loop}$, for sequences of various block lengths obtained for a block
copolymer DNA of length $N = 512$, with respect to $(g_0-g_c)^{\alpha}
\omega^{\beta}$, at three different force amplitudes $g_0 = 5.0$, 6.5,
and 8 in the low-frequency regime (i.e., $\omega \to 0$). In the above
expression we have subtracted the critical force $g_c$ needed to unzip
the block copolymer DNA for the static force case [$g_c(T = 4) =
3.0467.. $]. A similar type of scaling was found earlier for the
unzipping of a homopolymer DNA using Brownian
dynamics~\cite{Kumar2013,Kumar2016} and Monte Carlo
simulations~\cite{Kapri2014}. The exponents $\alpha$ and $\beta$ are,
however, found to be different for these studies. The earlier studies
on Brownian dynamics simulations suggested $\alpha = 1/2$ and $\beta =
1/2$~\cite{Kumar2013}, which were later modified to $\alpha = 0.33$ and
$\beta = 1/2$~\cite{Kumar2016}. On the other hand, the exponents
obtained for the Monte Carlo studies were $\alpha = 1$ and $\beta =
5/4$~\cite{Kapri2014}. In Figs.~\ref{fig:8}(a) and 8(b), we have plotted
the scaled data for three sequences $(A_4B_4)_{64}$, $(A_{64}B_{64})_4$,
and $(A_{128}B_{128})_2$ and the data for the opposite sequences
$(B_4A_4)_{64}$, $(B_{64}A_{64})_4$, and $(B_{128}A_{128})_2$,
respectively. For all these sequences, we obtain a nice collapse for
values $\alpha = 1.0 \pm 0.05$ and $\beta = 1.25 \pm 0.05$, the same as
the exponents obtained in earlier Monte Carlo studies for the unzipping
of a homopolymer DNA by a periodic force~\cite{Kapri2014}.

\section{Conclusions} {\label{sec:Summary} }

To summarize, in this paper we have studied the unzipping of a block
copolymer DNA subjected to a periodic force with amplitude $g_0$ and
frequency $\omega$ using Monte Carlo simulations. We obtained results
for the static force case and found that the equilibrium results do not
depend on the block copolymer DNA sequence and, the
temperature-dependent phase boundary $g_c(T)$, separating the zipped and
the unzipped phases, could be obtained by replacing the binding energy
in the exact expression previously obtained for the homopolymer DNA
case, by an effective average binding energy per block of the block
copolymer DNA sequence. For the dynamic case, the system, however, is
not in equilibrium, and results depend on the amplitude and frequency of
the periodic force as well as on the DNA sequence. We monitor the
separation between the strands of the block copolymer DNA as a function
of time at various frequencies and force amplitudes. The averaged
separation $\langle x \rangle$ plotted as a function of force value $g$
shows a hysteresis loop. The shape of the hysteresis loops is found to
be dependent on the frequency of the periodic force and the sequence of
block copolymer DNA. For sequences of shorter block lengths, e.g.,
$(A_4B_4)_{32}$ and $(B_4A_4)_{32}$, the loops are found to be the same,
with equal area, irrespective of periodic force acting on $A$- or
$B$-type base pairs at all frequencies. However, for longer block
lengths, e.g., $(A_{128}B_{128})_{1}$ and $(B_{128}A_{128})_{1}$, the
shape of the loops strongly depends on whether the periodic force is
applied on $A$- or $B$-type base pairs. We also obtain the area of the
hysteresis loops, $A_{loop}$ as a function of frequency $\omega$. The
resonance frequency, $\omega^{*}$, at which the loop area $A_{loop}$ is
maximum is higher for the sequence $(B_{128}A_{128})_{1}$. For
frequencies higher than $\omega^{*}$, the loop area for the sequence
$(B_{128}A_{128})_{1}$ is always more than that for the opposite
sequence $(A_{128}B_{128})_{1}$. Another difference is the oscillatory
behavior of the $A_{loop}$ seen for the sequence $(B_{128}A_{128})_{1}$,
whereas it is absent for the opposite sequence. For frequencies lower
than $\omega^{*}$, we find that the rate at which $A_{loop}$ decreases
with frequency also depends on the block length. The loop area for the
sequence $(B_{128}A_{128})_{1}$ is found to decrease much faster than
the sequence $(A_{128}B_{128})_{1}$. In the lower frequency regime
$A_{loop}$ scales as $A_{loop} \sim (g_0 - g_c)^{\alpha} \omega^{\beta}$
with exponents $\alpha = 1$ and $\beta = 5/4$ same as the exponents
obtained for periodic forcing of a homopolymer DNA studied
earlier~\cite{Kapri2014}, whereas in the higher frequencies, the loop
area $A_{loop}$ is found to scale with frequency as $A_{loop} \sim
1/\omega$~\cite{Mishra2013,Kapri2014}. The differences in exponents
observed in Brownian dynamics simulations and the current study requires
further investigation and will be the subject of a future study.

\section*{Acknowledgment}
We thank S. M. Bhattacharjee, A. Chaudhuri and S. Kalyan
for comments and discussions.

\end{document}